\documentstyle[12pt]{article}

\newcommand{\vev}[1]{\langle #1 \rangle}                 
\newcommand{\com}[2]{\left[ #1,#2 \right]}               
\newcommand{\resetcounter}{\setcounter{equation}{0}}     
\newcommand{\lagrange}{{\cal L}}

\newcommand{\D}  {{\cal D}}               
\newcommand{\EPS}{{\cal E}}               
\newcommand{\R}  {{\cal R}}               
\newcommand{\W}  {{\cal W}}               
\newcommand{\OA}  {\Omega}                
\newcommand{\Si}  {\Sigma}                
\newcommand{\ga}{\gamma}                  
\newcommand{\GA}{\Gamma}                  
\newcommand{\T}  {\theta}

\newcommand{\A}  {\alpha}
\newcommand{\B}  {\beta}
\newcommand{\E}  {\varepsilon}

\newcommand{\Om}  {\omega}

\newcommand{\Sm} {\sigma}
\newcommand{\LAB} {\Lambda}               

\newcommand{\C} {\chi}

\begin{document}

\thispagestyle{empty}
\begin{titlepage}
\begin{flushright}
HUB--IEP--94/33 \\
hep-th/9412079 \\
December 1994
\end{flushright}
\vspace{0.3cm}
\begin{center}
\Large \bf Gaugino Condensation
           \\ in the Chiral and Linear Representation
           \\ of the Dilaton
\end{center}
\vspace{0.5cm}
\begin{center}
Ingo \ Gaida$^{\hbox{\footnotesize{1,2}}}$ and
Dieter \ L\"ust$^{\hbox{\footnotesize{1}}}$,  \\
{\sl Institut f\"ur Physik, Humboldt--Universit\"at,\\
 Invalidenstrasse 110, D--10115 Berlin, Germany}
\end{center}
\vspace{0.6cm}

\begin{abstract}
\noindent
String effective theories with N=1 supersymmetry in four dimensions
are subject of the discussion.
Gaugino condensation in the chiral representation
of the dilaton is reviewed in the truncated formalism in the
$U_{K}(1)$-superspace. Using the supersymmetric duality of the
dilaton the same investigation is made in the linear
representation of the dilaton. We show that for the
simple case of one gaugino condensate the results concerning
supersymmtry breaking are  independent of the
representation of the dilaton.
\end{abstract}

\vspace{0.3cm}
\footnotetext[1]{E-MAIL: gaida@qft2.physik.hu-berlin.de,
                         luest@qft1.physik.hu-berlin.de.}
\footnotetext[2]{Supported by Cusanuswerk}
\vfill
\end{titlepage}


\setcounter{page}{1}

%
%
%

\section{Introduction}
\resetcounter

String effective theories with N=1 supersymmetry in 4 dimensions
are subject of the discussion.
These theories are effective in the sense, that they are
low-energy limits of a given higher dimensional string theory
after dimensional reduction and integrating out all heavy modes.
At tree level the gauge coupling constant can be expressed
by the vacuum expectation value of the dilaton
superfield
$S$: $g^{2} = 2 \ \vev{S+ \bar S}^{-1}$.
Throughout this text $S+ \bar S$ will be denoted as
the chiral representation of the dilaton.

It has been shown
that there exists a supersymmetric legendre transformation
called {\em supersymmetric duality}, which transforms
$S+ \bar S$ into a linear superfield $L$
[\ref{linear},\ref{Grimm1},\ref{Chern_Simons},
\ref{ovrut_schwiebert},\ref{Cardoso},\ref{derendinger},\ref{derque}],
where $L$ will be called the linear representation of
the dilaton. From the point of view
of  string theory this is very attractive,
because the linear superfield contains a real scalar, denoted
as the dilaton $C$ here, and an antisymmetric tensor. This
field content naturally exists together with the graviton
at the massless level of string theories. The corresponding
vertex operators are

\begin{eqnarray}
V &=& \E_{\mu \nu} \ \partial X^{\mu} \ \partial X^{\nu}
      \ e^{ikX}
\nonumber\\
\end{eqnarray}

with $ \E_{\mu \nu} = \E_{\nu \mu } $ for the graviton,
$ \E_{\mu \nu} = - \E_{\nu \mu } $ for the antisymmetric tensor
and
$ \E_{\mu \nu} = \E_{\nu \nu } $ for the dilaton.

Since one integrates out only the massive states to go to
the effective theory, these fields appear in the low energy supergravity action
of four-dimensional, $N=1$ supersymmetric heterotic strings.
Due to phenomenological reasons
 $N=1$ supersymmetry   must be broken.
 Non-perturbative condensation [\ref{gaugino}]
of gauginos provides a promising mechanism
for spontaneous supersymmetry breaking
at the level of the effective supergravity action and has been
studied extensively [\ref{dual},\ref{filq},\ref{fmtv}]
in the pure chiral formulation
of global and local supersymmtry.

According to Witten's index theorem [\ref{witten_index}]
global supersymmetry is not broken in a pure Super-Yang-Mills
model or a Super-Yang-Mills model coupled to real matter
representation, at least in the case of massive matter fields.
This leads to the so called {\em truncated formalism} in studying
gaugino condensation and will be very powerful in the following
discussion.

Since, as already mentioned, the linear representation of the dilaton plus
antisymmetric tensor field appears to be more natural in string theory,
it is an important task to see whether supersymmetry breaking by
gaugino condensation can be also consistently formulated within the linear
formalism. This topic was essentially not yet discussed in the  literature
(for some interesting considerations see [\ref{derque}]),
and we try to make a
step to fill this gap.
An important aspect within this framework is that
the linear formalism automatically preserves an
$U(1)_{PQ}$ symmetry, namely shifts in the antisymmetric tensor field of the
form $b_{mn} \rightarrow b_{mn} + \partial_{m} b_{n} - \partial_{n} b_{m}$
with $b_{m}$ arbitrary.
Thus describing the gaugino condensation in the linear formalism implicitely
assumes that the $U(1)_{PQ}$ symmetry is not broken by the non-perturbative
dynamics. (Perturbatively, the $U(1)_{PQ}$ holds in any case.)
We will show in the following that the formation of one gaugino condensate
in one hidden gauge sector can be consistently formulated within the linear
formalism. The case of several gaugino condensates is planned
to be investigated in the future.

\vspace{0.5cm}

The paper is organized as follows:
After definition of the field content and
discussion of the symmetries, a short introduction
in the $U_{K}(1)$-superspace is given.
Afterwards gaugino
condensation is studied in the chiral and
the linear representation of the dilaton. The
supersymmetric duality is also investigated
in the Lorentz-superspace, where some subtleties
occur.


\section{Field Content and Symmetries}
\resetcounter

$N=1$ supersymmetric theories can be easily defined in superspace
[\ref{wess_and_bagger},\ref{Grimm1},\ref{Grimm_2}]. One of
the basic objects in any superspace formulation are
chiral superfields $\Si$, because their
lowest components are scalar fields parametrizing
a K\"ahler manifold [\ref{zumino}].
These chiral superfields obey the constraint

\begin{eqnarray}
\bar\D^{\dot\A} \ \Si &=& 0
\hspace{2cm}
\D_{\A} \ \bar\Si = 0
\nonumber\\
\end{eqnarray}

and are defined at component level\footnote{We use the usual
superspace notations $\int \equiv \int d^{4}\theta$ and
$X_{|} \equiv X_{| \underline\theta = 0}$.
}

\begin{eqnarray}
\Si_{|} &=& A(x)
\hspace{1cm}
\D_{\A} \Si_{|} = \sqrt{2} \ \C_{\A}(x)
\hspace{1cm}
\D^{2} \Si_{|} = - 4 \ F(x).
\nonumber\\
\end{eqnarray}

The Yang Mills multiplet $W_{\A}^{ \ (r)}$ with the index
$r$ belonging to the internal gauge group
is defined as

\begin{eqnarray}
W_{\A |}^{ \ (r)} &=& -i \ \lambda_{\A}^{ \ (r)}(x)
\hspace{2cm}
\D_{\B} W_{\A |}^{ \ (r)}  = - \E_{\B\A} D^{(r)}(x)
  -i \ (\Sm^{mn}\E)_{\B\A} \ f_{mn}^{ \ \ (r)}(x)
\nonumber\\
\end{eqnarray}

and the field strength reads

\begin{eqnarray}
f_{mn}^{ \ \ (r)} &=& \partial_{m} a_{n}^{ \ (r)}
                     -  \partial_{n} a_{m}^{ \ (r)}
                     +  a_{m}^{ \ (p)} a_{n}^{ \ (q)}
                        \ c_{(p)(q)}^{ \ \ \ \ \ (r)}
\nonumber\\
\end{eqnarray}

with the generators of the internal gauge group
satisfying

\begin{eqnarray}
\com{T_{(r)}}{T_{(s)}} &=& i \ \ c_{(r)(s)}^{ \ \ \ \ \ (q)} \ T_{(q)}.
\nonumber\\
\end{eqnarray}

By constructing the chiral density
$\EPS = e + i e \T \Sm^{a} \bar\psi_{a} - e \T^{2} (\bar M +
        \bar\psi_{a}\bar\Sm^{ab}\bar\psi_{b})$
one finds for the supergravity sector
the graviton, the gravitino and
two auxiliary fields [\ref{wess_and_bagger}]

\begin{eqnarray}
E_{m \ \ |}^{ \ a}  = e_{m}^{ \ a}(x)
\hspace{1cm}
E_{m \ \ |}^{ \ \A} = \frac{1}{2} \psi_{m}^{ \ \A}(x)
\hspace{1cm}
R_{|}  = -  \frac{1}{6} M(x)
\hspace{1cm}
G_{a |}  = - \frac{1}{3} b_{a}(x).
\nonumber\\
\end{eqnarray}

The linear multiplet $L$ is the difference of the Chern-Simons
superfield $\OA$ [\ref{Chern_Simons},\ref{derendinger},\ref{Cardoso},
\ref{ovrut_schwiebert}]
and the real linear multiplet $l$
and satisfies
the following constraint:

\begin{eqnarray}
\label{linear_multiplet_constraint}
 (\bar\D^{2} - 8 R) \ L &=& -2 \ \mbox{tr} \ W^{\A} W_{\A}
\hspace{2cm}
L = l - \OA = L^{+}
\nonumber\\
\end{eqnarray}

The real linear multiplet $l$ obeys $(\bar\D^{2} - 8 R) \ l =  0$
and therefore  $(\bar\D^{2} - 8 R) \ \OA = 2 \ \mbox{tr} \ W^{\A} W_{\A} $.
The linear multiplet $L$ contains a real scalar $C$, which is called
dilaton in this framework,
its supersymmetric partner, the dilatino $\varphi_{\A}$,
an antisymmetric tensor $b_{mn}$ and the
Yang-Mills Chern-Simons form
$\Om_{3Ynml} = - \mbox{tr} (a_{[l}\partial_{m}a_{n]}
               - \frac{2i}{3} \ a_{[l} a_{m} a_{n]} )$
:

\begin{eqnarray}
\mbox{ln} L_{|} &=& C(x)
\nonumber\\
\nonumber\\
\D_{\A} \ \mbox{ln} L_{|}  &=& \varphi_{\A}(x)
\nonumber\\
\nonumber\\
\com{\D_{\A}}{\bar\D_{\dot\A}} L_{|}
&=& - \frac{4}{3}  e^{C} b_{\A \dot\A}
    + 4 \ \mbox{tr} \ \lambda_{\A} \bar\lambda_{\dot\A}
   + \Sm_{k \A \dot\A}
   \left \{
        \E^{klmn} (\partial_{n} b_{ml} - \frac{1}{3} \ \Om_{3Ynml}
   \right .
\nonumber\\ & &
   \left .
        + i \ e^{C} \psi_{n} \Sm_{m} \bar\psi_{l})
        +2i \ e^{C} ( \psi_{m} \Sm^{mk} \varphi -
                      \bar\psi_{m} \bar\Sm^{mk} \bar\varphi)
   \right \}
\nonumber\\
\end{eqnarray}

The supersymmetric duality of the dilaton will be discussed in this
paper and the duality transformed linear multiplet will
be denoted as $S_{R} = S + \bar S$ with $S$ being a chiral
superfield. Inspired by [\ref{dim_red}] one defines

\begin{eqnarray}
S_{|} &=& (e^{-C} \ + \ i \ a )(x)
\hspace{1cm}
\D_{\A} S_{|} = \sqrt{2} \ \rho_{\A}(x)
\hspace{1cm}
\D^{2}  S_{|} = - 4 \ f(x)
\nonumber\\
\end{eqnarray}

Effective superstring models do not only have the symmetries
of usual supergravity theories, namely K\"ahler symmetry
[\ref{Grimm1}]
or super-Weyl-K\"ahler symmetry [\ref{wess_and_bagger}],
but also target space modular symmetries (for a review see [\ref{gpr}]) induced
by the target space duality group. For orbifold compactifications
the target space duality group is often given by the modular group
$PSL(2,{\bf Z})$,
acting on one chiral field $T$ as\begin{eqnarray}
\label{modular_transformation_of_moduli}
 T^{\prime}   &=& \frac{a \ T - i \ b}{ i \ c \ T + d } \hspace{1cm}
       ad - bc     =  1 \hspace{1cm}  a,b,c,d \in {\bf Z}
\nonumber\\
\end{eqnarray}
where $T$ corresponds to an internal, overall modulus: $T=R^2+iB$.
We just consider this case for simplicity. Then the effective
supergravity action has to be target space modular invariant [\ref{FLST}].

\vspace{0.5cm}
The target space modular transformations act as K\"ahler transformations
on the K\"ahler potential.
K\"ahler transformations are harmonic transformations of the
K\"ahler potential of the form

\begin{eqnarray}
\label{Kahler_transformation}
     K^{\prime} \ (\Si, \bar\Si) &=&
     K (\Si, \bar\Si) + \ F(\Si) +  \ \bar F(\bar\Si),
\nonumber\\
\end{eqnarray}

where $F(\Si)$ are arbitrary holomorphic functions.
The superpotential $\Om(\Si)$ must transform in the following way
under K\"ahler transformations:

\begin{eqnarray}
\label{superpotential_Kahler_transformation}
     \Om^{\prime} \ (\Si) &=&  \Om(\Si) \ e^{- F(\Si)}
\nonumber\\
\end{eqnarray}

The (super-Weyl-) K\"ahler and the modular symmetry can be
characterized by
weights. With respect to one of these symmetries
the weights add up and an invariant expression has
vanishing weight:

\begin{eqnarray}
\label{weight_rule}
 \W(X \ Y)     &=&  \W(X) + \W(Y)  \hspace{1cm} ,  \hspace{1cm}
    X^{\prime}  =  X   \ \Leftrightarrow \  \W(X) = 0
\nonumber\\
\end{eqnarray}


\section{The $U_{K}(1)$-Superspace}
\resetcounter

In the $U_{K}(1)$-superspace formulation [\ref{Grimm1},\ref{Grimm_2}]
all the weights
of these different symmetries can be characterized
by K\"ahler weights only. Furthermore the $U_{K}(1)$-superspace
is manifestly Einstein normalized. This issue will
be very powerful in the discussion of the
supersymmetric duality of the dilaton.

\vspace{1cm}

The structure group of $U_{K}(1)$-superspace is
$SL(2,{\bf C}) \times  U_{K}(1)$ and the corresponding
two Lie algebra valued connections are

\begin{eqnarray}
\label{connection_structure_group}
   \phi_{B}^{ \ \ A}  = dz^{M} \   \phi_{MB}^{ \ \ \ \ A}
\hspace{1cm} \mbox{and}  \hspace{1cm}
   A  = dz^{M} \ A_{M}
\nonumber\\
\end{eqnarray}

The  $U_{K}(1)$ connection $A$ is a composite
gauge connection defined as

\begin{eqnarray}
\label{u1_connection}
   A_{\A}          = \frac{1}{4} \ \D_{\A} K
\hspace{1cm}
   \bar A^{\dot\A} = -\frac{1}{4} \ \bar\D^{\dot\A} K
\hspace{1cm}
   A_{\A\dot\A}    = \frac{3i}{2} \ G_{\A\dot\A}
                      - \frac{i}{8} \ \com{\D_{\A}}{\bar\D_{\dot\A}} K,
\nonumber\\
\end{eqnarray}

and $K$ is the K\"ahler potential
defining a K\"ahler manifold.
The corresponding K\"ahler-form, also denoted as $K$ for simplicity,
obeys the K\"ahler condition [\ref{zumino}]:

\begin{eqnarray}
\label{kahler_condition}
 K = K^{+} = \frac{i}{2} \ dA^{i} d{\bar A}^{\bar j} \ g_{i {\bar j} }
 \hspace{1cm}  \mbox{and} \hspace{1cm}
 dK = 0,
\nonumber\\
\end{eqnarray}

where $g_{i {\bar j}} =
  \partial_{i} \ \partial_{\bar j} \ K(A, {\bar A})$.

Introducing an internal gauge group with
connection $ h_{a}^{ \ b} = dz^{M} \ h_{Ma}^{ \ \ \ b} $
one can solve the Bianchi-identities subject
to a set of constraints [\ref{martin_mueller}].
It turns out, that the basic objects of the theory
are the following superfields with
K\"ahler weights $\W_{K}$

\vspace{1cm}

\begin{center}
\begin{tabular}{|c||c|c|c|c|c|}
\hline
& & & & & \\
Superfield & $R$ & $W_{\A\B\ga},W_{\A},X_{\A}$ & $G_{a}$
                 & $\bar W_{\dot\A\dot\B\dot\ga}
                    ,\bar W_{\dot\A}
                    ,\bar X_{\dot\A}$
                 & $R^{+}$
\\
& & & & & \\
\hline
& & & & & \\
$\W_{K}$ &$ 2 $&$ 1$ &$ 0$ &$ -1$ &$ -2$ \\
& & & & & \\
\hline
\end{tabular}
\end{center}

\vspace{1cm}

Here $X_{\A}$ is the field strength of the $U_{K}(1)$
connection with

\begin{eqnarray}
\label{u(1)_field_strenth}
X_{\A} &=& \D_{\A} R - \bar\D^{\dot\A} G_{\A \dot\A}
        =  - \frac{1}{8} (\bar\D^{2} - 8R) \ \D_{\A} K
\nonumber\\
\bar X^{\dot\A} &=& \bar\D^{\dot\A} R - \D_{\A} G^{\A \dot\A}
        =  - \frac{1}{8} (\D^{2} - 8R^{+}) \ \bar\D^{\dot\A} K
\nonumber\\
\end{eqnarray}

The resulting action is invariant under K\"ahler,
gauge and Lorentz transformations [\ref{Grimm1},\ref{Cardoso}].
Superfields $Z$ with K\"ahler weight $\W_{K}(Z) = \gamma$
transform under K\"ahler transformation as

\begin{eqnarray}
\label{kaehler_trafo_general}
 Z \  \ \rightarrow \ \    Z \ e^{-i \gamma Im F(\Si)/2}
\nonumber\\
\end{eqnarray}

and the covariant derivative is given as

\begin{eqnarray}
\label{u1_covariant_derivative}
 \D_{B} Z &=& E_{B}^{ \ \ M} \partial_{M} Z + \gamma A_{B} Z.
\nonumber\\
\end{eqnarray}

To build a model it is necessary to know about three functions
[\ref{sugra_1}],
namely the K\"ahler potential $K$, the superpotential $\Om$ and the
gauge kinetic function $f_{ A B}$. We now consider the case of one
gaugino condensate in a pure Yang-Mills hidden gauge sector with gauge group
$G$. The corresponding $N=1$ $\beta$-function coefficient is denoted by
$b$ ($b=-3N$ for $G=SU(N)$).
The induced effective superpotential
dictated by Ward identities and
modular invariance, is non perturbative and reads
[\ref{taylor1}]

\begin{eqnarray}
\label{effective_superpotential}
  \Om(\Si) &=& \frac{1}{8 \pi^{2}} \ Y^{3} e^{-K/2} \
             \mbox{ln}
             \left \{
                     c^{-2b}e^{8\pi^2f(\Si)} \ Y^{-b} \ e^{Kb/6}
             \right \},
\end{eqnarray}

where $f_{AB} \sim \delta_{AB} \ f(\Si)$.
The light degrees of freedom are represented by gauge-singlet
composite superfields: $Y^{3} \sim W^{\A} W_{\A}$ with the
K\"ahler weight $\W_{K}(Y) = 2/3$.
The lagrangian in the $U_{K}(1)$-superspace formulation is given in
[\ref{Grimm_2}]

\begin{eqnarray}
\label{U1_lagrangian_general}
 \lagrange &=&  \lagrange_{matter} +
                \lagrange_{YM} +
                \lagrange_{pot}
\end{eqnarray}

with

\begin{eqnarray}
\label{U1_lagrangian}
 \lagrange_{matter}&=&  - 3 \int \ E[K]
\nonumber\\
 \lagrange_{pot}   &=&\frac{1}{2} \int \frac{E}{R} \ e^{K/2} \ \Om(\Si)
                      + h.c.
\nonumber\\
 \lagrange_{YM}    &=& \frac{1}{2} \int \frac{E}{R} \ W^{A} \ f_{AB} \ W^{B}
                       + h.c.
\nonumber\\
\end{eqnarray}

\section{Gaugino Condensation}
\resetcounter

Now gaugino condensation can be studied in
the framework of $U_{K}(1)$-superspace.
 Everything is worked out in the {\em truncated formalism}.


\subsection{The Chiral Representation of the Dilaton}

The tree-level K\"ahler potential under discussion will be

\begin{eqnarray}
\label{matter_kaehler_potential}
 K  &=& -4n \ \mbox{ln} (S + \bar S) -
           \ 3 \ \mbox{ln} (T + \bar T)
\end{eqnarray}

and the gauge kinetic function reads at tree level plus one loop [\ref{DKL}]
(we neglect
effects from possible Green-Schwarz terms, i.e. we assume that $S$ is
invariant under target space modular transformations)

\begin{eqnarray}
\label{matter_kaehler_potential}
 f_{AB} &=&  n \ \delta_{AB} \
                         \left(
                         S \ -{b\over 8\pi^2}\ln\eta(T)^2
                         \right)
\end{eqnarray}
$\eta(T)$ is the well-known Dedekind function.
So the gauge kinetic function leads to an effective superpotential [\ref{fmtv}]

\begin{eqnarray}
\label{effective_superpotential}
  \Om(\Si) &=& \frac{1}{8 \pi^{2}} \ Y^{3} e^{-K/2} \
             \mbox{ln}
             \left \{
                     (c \eta(T))^{-2b} \ Y^{-b} \
                      e^{8 \pi^{2} S}  \ e^{Kb/6}
             \right \}
\end{eqnarray}

with $Y^{3} = n \ W^{\A}W_{\A}$.
Note that in the $U_{K}(1)$-superspace $Y$ transforms under
K\"ahler transformations as
$Y^{\prime} = Y \ e^{-(F(\Si) - \bar F(\bar\Si))/6}$.
But the combination $Y \ e^{-K/6}$ transforms purely chiral:

\begin{eqnarray}
(Y \ e^{-K/6})^{\prime} &=& (Y \ e^{-K/6}) \ e^{-F(\Si)/3}
\end{eqnarray}

If global supersymmetry is not broken, one can eliminate the
condensate $Y$ and go to the {\em truncated} lagrangian.
The condition for unbroken global supersymmetry reads

\begin{eqnarray}
\label{unbroken_superpotential_constraint}
 \frac{\partial \Om}{\partial Y} &=& 0
\end{eqnarray}

and yields the equations of motion for the composite field
$Y$.

\begin{eqnarray}
\label{EOM_for_condensates}
  Y = 0
\hspace{1cm} , \hspace{1cm}
  \frac{1}{8 \pi^{2}} \ Y^{3} = (8 \pi^{2} e)^{-1} \
                                       (c \eta(T))^{-6} \
                                   e^{ 24 \pi^{2} S/b} \ e^{K/2}
\end{eqnarray}

The first equation of motion is just the trivial minimum corresponding
to $V=0$. For gaugino condensation one must find an energetically
favoured minimum compared to this trivial one.
Note that the second equation is equivalent
to the statement, that the superpotential is chiral. It vanishes in
the flat limit, where $e^{K} \rightarrow 1$ in
(\ref{effective_superpotential}).
Therefore the second equation of motion {\em must} hold and will be
studied further.
The truncated superpotential reads now [\ref{filq}]

\begin{eqnarray}
\label{truncated_superpotential}
  \Om(S, T) &=& g(S) \ h(T)
\end{eqnarray}

with the definition

\begin{eqnarray}
\label{def_truncated_superpotential}
 g(S) &=&  \frac{b}{3} \ e^{- \B  S}
\\
 h(T) &=& (8 \pi^{2}e)^{-1} (c \eta(T))^{-6}
\\
 \B   &=& -24 \pi^{2}/b
\end{eqnarray}

As usual one performs a K\"ahler transformation to
go to the G-function:

with $ G  = K + \mbox{ln} |\Om|^{2} $
and $ \eta^{-6 \ \prime} = \eta^{-6} \Om^{-1} $
the lagrangian reads

\begin{eqnarray}
\label{truncated_chiral_u1_lagrangian_1}
  \lagrange &=& - 3 \int \ E[G] +
                \left(
                 \frac{1}{2} \int \ \frac{E}{R} \ e^{G/2} + h.c.
                \right) ,
\nonumber\\
\end{eqnarray}

if - as considered here - the superpotential is non-singular.
Going to components the potential reads
in general [\ref{sugra_1}]

\begin{eqnarray}
\label{general_potential}
 V  &=& e^{G} \ (G_{i} \ G^{i \bar j} G_{\bar j} \ - \  3).
\nonumber\\
\end{eqnarray}

If one introduces now the Eisenstein function
$ \hat{G}_{2}(T) = G_{2}(T) \ - \ \frac{2\pi}{T_{R}} $
and defines
$k(S + \bar S) =  (\frac{\B}{2\sqrt{n}} S_{R} + 2 \sqrt{n})^{2}$
one finds the effective potential in the chiral representation
of the dilaton

\begin{eqnarray}
\label{modular_inv_truncated_potential}
 V (S,T) &=&
      \frac{ | \Om |^{2} }{S_{R} \ T_{R}^{3}} \
      \left \{
              \frac{3T^{2}_{R}}{4 \pi^{2}} \ |\hat{G}_{2}(T)|^{2}
              + k(S + \bar S) - 3
      \right \}
\end{eqnarray}

Note that the potential is positive for a constant superpotential,
say $\Om = 1$, if the K\"ahler potential is of the
considered no-scale type.

\subsection{Duality Transformation}

The supersymmetric duality transformation amounts to a
legendre transformation, which transforms the dilaton
superfield $S+\bar S$ into the linear multiplet $L$
[\ref{Chern_Simons},\ref{Grimm1}].

\begin{eqnarray}
 \lagrange &=&  \lagrange_{matter} +  \lagrange_{YM} + \lagrange_{pot}
\nonumber\\
&=&
 - 3 \int \ E[G] +  \lagrange_{pot}
 + ( \frac{n}{2} \int \ \frac{E}{R} \ S \ W^{\A} W_{\A} \ + \ h.c. )
\nonumber\\
&=&
  - 3 \int \ E \ \left (
                        1 + \frac{2n}{3} (S + \bar S) \ \OA
                 \right )
  + \lagrange_{pot}
\end{eqnarray}

Now one adds a lagrange multiplier $\lagrange_{lm}$
and sets $S + \bar S = U$,
where $U$ is unconstrained. The lagrange multiplier contains
the unconstrained field $U$ and the real linear multiplet $l$.
The whole system adds up to an unconstrained lagrangian $\lagrange_{u}$.

\begin{eqnarray}
\lagrange_{u} &=&
    - 3 \ \int \ E \
    \left (
           1 - \frac{2n}{3} \ U \ L
    \right )
    + \lagrange_{pot}
\end{eqnarray}

Variation with respect to $l$ yields the old theory in the
chiral representation of the dilaton.
Variation with respect to $U$ yields the linear representation of the
dilaton. The following variations in $U_{K}(1)$-superspace must be used:

\begin{eqnarray}
 \delta_{U} E  &=& -\frac{1}{3} \  \frac{\partial K}{\partial U} \
                   \delta U \ E
\nonumber\\
 \delta_{U} R  &=&  \frac{1}{6} \  \frac{\partial K}{\partial U} \
                   \delta U \ R
\nonumber\\
 \delta_{U} L &=& \frac{1}{3} \  \frac{\partial K}{\partial U} \
                   \delta U \ L
\end{eqnarray}

Note that $\lagrange_{pot}$ is invariant under variations with respect
to $U$ because $\Om \neq \Om(U)$.
The other part of the unconstrained lagrangian
yields

\begin{eqnarray}
 \frac{\partial K}{\partial U} (1-\frac{2n}{3}\ U \ L)
 + 2n \ L (1+\frac{1}{3} \frac{\partial K}{\partial U} \ U)
 &=& 0
\end{eqnarray}

So one finds with $K(U) = -4n \ \mbox{ln} U$
the {\em duality relation}

\begin{eqnarray}
\label{duality_relation}
 L  &=& \frac{2}{U}.
\end{eqnarray}

Inserting this relation, one ends up with the new dual lagrangian.

\begin{eqnarray}
\label{dual_action}
 \lagrange_{dual} &=&
 (4n - 3) \int E[\GA]
 + \left(
      \frac{1}{2} \int  \frac{E}{R} \ e^{\GA/2} + h.c.
   \right)
\nonumber\\
\end{eqnarray}

with the dual G-function $\GA$ given by

\begin{eqnarray}
\label{dual_g_function}
 \GA &=& 4n \ \mbox{ln} \ L -  4n \ \mbox{ln} \ 2
         \ - 3 \ \mbox{ln} (T + \bar T) +  \mbox{ln} |\Om|^{2}
\nonumber\\
\end{eqnarray}

Note that the duality transformation has {\em eaten} the
Yang-Mills tree-level action of the chiral
representation of the dilaton.
This part of the action will not enter an effective
superpotential anymore. In the following it will be shown,
that also in the potential this part is absent.
That is to say, the characteristic term $e^{- \B S}$ vanishes
in the linear representation of the dilaton.

In the dual theory one needs only to know two functions,
namely the dual K\"ahler potential
$\tilde K_{dual} =4n \ \mbox{ln} L  -  4n \ \mbox{ln} \ 2
\ - 3 \ \mbox{ln} (T + \bar T)$ and the
superpotential $\Om$. The price one has to pay is,
that the dual K\"ahler potential $\tilde K_{dual}$ is ill defined
in the sense, that it is not a K\"ahler potential defining a
K\"ahler manifold. However this can be healed by
demanding, that the well-defined dual K\"ahler
potential $K_{dual}$ does not depend on the linear multiplet L.
That is to say $K_{dual} = \tilde K_{dual} - f(L)$,
with $f(L) = 4n \ \mbox{ln} (L/2)$ in this specific considerations.
So by identifying the well defined K\"ahler potential one
ends up again with three functions, one has to know
for model building. But in this context the question arises,
if the dual K\"ahler potential can always be decomposed
in such a way.

\subsection{Linear Representation of the Dilaton}

In the linear representation of the dilaton\footnote{The index
{\em dual} will be dropped from now on.} the
effective superpotential is given by
$\Om(\Si) = \frac{1}{8 \pi^{2}} \ Y^{3} e^{-\tilde K/2} \
             \mbox{ln}
             \left (
                     (c \eta(T))^{-2b} \ Y^{-b}  \ e^{\tilde Kb/6}
             \right )$
only. Again one eliminates the condensate and finds the
truncated superpotential

\begin{eqnarray}
\label{linear_truncated_superpotential}
  \Om(T) &=& g \ h(T)
\end{eqnarray}

with the definition

\begin{eqnarray}
\label{def_truncated_superpotential}
 g &=&  \frac{b}{3}
\hspace{1cm} \mbox{and}  \hspace{1cm}
 h(T) = (8 \pi^{2}e)^{-1} (c \eta(T))^{-6} .
\end{eqnarray}

The equation of motion for the condensate reads

\begin{eqnarray}
  \frac{1}{8 \pi^{2}} \ Y^{3} &=& (8 \pi^{2} e)^{-1} \
                                        (c \eta(T))^{-6} \
                                     \ e^{\tilde K/2}
\end{eqnarray}

Again one performs a K\"ahler transformation to
go to the $\GA$-function:
With $\GA  = \tilde K + \mbox{ln} |\Om|^{2}$
and  $\eta^{-6 \ \prime} = \eta^{-6} \Om^{-1}$
the whole lagrangian is given by (\ref{dual_action}).
$\lagrange_{pot}$ transforms in the following way
under this K\"ahler transformation:

\begin{eqnarray}
\label{truncated_linear_u1_superpotential}
  \lagrange_{pot}(\tilde K,\Om)
       = \frac{1}{2} \int \ \frac{E}{R} \ e^{\tilde K/2} \ \Om(T) + h.c.
  \hspace{0.5 cm} \rightarrow \hspace{0.5 cm}
 \lagrange_{pot}(\GA)
       = \frac{1}{2} \int \ \frac{E}{R} \ e^{\GA/2} + h.c.
\nonumber\\
\end{eqnarray}

The equation of motion for the condensate also transforms under
this K\"ahler transformation.
At component level one finds:

\begin{eqnarray}
\label{gaugino_constraint}
  tr \ \lambda^{\A} \lambda_{\A} &=& \frac{\B}{n} \ e^{\GA/2}
\end{eqnarray}

At this point one has to ask for the potential.
This time the potential is given by two parts. One is
the usual scalar part given by eliminating the auxiliary fields
$M$ and $F^{i}$. The second part is given by contributions, where the
dilaton couples to the gaugino condensate only.

The lagrangian considered here is evaluated at component level
in [\ref{Grimm1}].
The part of the lagrangian containing the auxiliary fields
reads

\begin{eqnarray}
 \lagrange_{aux}/e &=& \frac{4n-3}{9} M \bar M
                     + g_{i \bar j} F^{i} {\bar F}^{\bar j}
                     + e^{\GA/2} \left (
                                  \frac{4n-3}{3}(M+\bar M)
                                  +  F^{i} G_{i}
                                  + \bar F^{\bar j} \bar G_{\bar j}
                                 \right ),
\nonumber\\
\end{eqnarray}

where G is the well-defined dual G-function:
$ G \equiv G_{dual} = K_{dual} + ln |\Om|^{2}$.
So the scalar potential from this part after elimination of the
auxiliary fields is

\begin{eqnarray}
\label{linear_potential_1}
 V  &=& e^{\GA} \ (G_{i} \ G^{i \bar j} G_{\bar j}  + 4n - 3 )
\nonumber\\
\end{eqnarray}

and reduces to the known potential in the chiral limit $n=0$.
It is positive again for the considered
no-scale K\"ahler potential $\tilde K$ and $\Om=1$ - in analogy
to the chiral representation of the dilaton.

The second part of the potential is the sum of all monomials
coupling the dilaton $C$ only to the condensate  tr $\lambda^2$:

\begin{eqnarray}
 \lagrange(C,\lambda^2 )/e &=&
    -2n \ e^{-C} \ e^{\GA/2}
    (\mbox{tr}\lambda^2 + \mbox{tr}\bar \lambda^2 )
    -n \ e^{-2C} \ \mbox{tr}\lambda^2 \  \mbox{tr}\bar \lambda^2
\nonumber\\
\end{eqnarray}

Eliminating the condensate by the equation of motion
(\ref{gaugino_constraint}) one finds

\begin{eqnarray}
\label{dilaton_part}
 V_{2}(C,T) &=&   \ e^{\GA} \
                      \left \{
                              (\frac{\B}{\sqrt{n}} \ e^{-C} + 2\sqrt{n})^2
                                - 4n
                      \right \}
 \nonumber\\
\end{eqnarray}

The sum of (\ref{linear_potential_1}) and (\ref{dilaton_part})
gives the full scalar potential in the linear representation
of the dilaton, if gaugino condensation takes place.
In analogy to the chiral case one defines
$ k(C) = (\frac{\B}{\sqrt{n}} \ e^{-C} + 2\sqrt{n})^2 $
and finds for general n the identity  $k(C) = k(S + \bar S)$
and

\begin{eqnarray}
\label{full_linear_potential}
 V(C,T) &=&  e^{\GA} \ \left \{
                               G_{i} \ G^{i \bar j} G_{\bar j}
                               + k(C) - 3
                        \right \}
 \nonumber\\
\end{eqnarray}

And for n=1/4 this reduces to

\begin{eqnarray}
\label{full_linear_potential}
 V(C,T) &=&
      \frac{ e^{C} }{2 T_{R}^{3}} \ | \Om |^{2}
      \left \{
              \frac{3T^{2}_{R}}{4 \pi^{2}} \ |\hat{G}_{2}|^{2}
              + k(C)  - 3
      \right \},
\end{eqnarray}

where $\Om$ is given by (\ref{linear_truncated_superpotential}).

The parameter $n$ must be equal to $1/4$ to lead to
a {\em good} duality preserving the canonical form
for the Yang Mills action at component level
[\ref{Grimm1}].
The mechanism of supersymmetry breaking by this effective
potential, discussed in the chiral representation of the
dilaton in [\ref{filq}],
is still valid, although the potential looks slightly
different.

\section{Connection to Lorentz-Superspace}
\resetcounter


In the Lorentz-Superspace of [\ref{wess_and_bagger}]
the considered lagrangian looks like

\begin{eqnarray}
\label{lorentz_chiral_action}
\lagrange
= - 3 \int \EPS
    \left \{
     e^{-K/3} + \frac{2n}{3} (S + \bar S) \OA
    \right \} + ( \frac{1}{2} \int \frac{\EPS}{R} \ \Om(\Si) + h.c. )
\end{eqnarray}

This lagrangian is invariant under the super-Weyl-K\"ahler
symmetry. The torsion constraints of Lorentz superspace
[\ref{martin_mueller},\ref{wess_and_bagger}]
are invariant under super-Weyl transformations [\ref{howe_tucker}].
The supervielbein transforms under super-Weyl transformations
with chiral superfields $\LAB$
as follows:

\begin{eqnarray}
\EPS_{M}^{ \ \ a} &\rightarrow& e^{\bar\LAB + \LAB} \ \EPS_{M}^{ \ \ a}
\nonumber\\
\EPS_{M}^{ \ \ \A} &\rightarrow& e^{2\bar\LAB - \LAB} \
                                 \left (
                                   \EPS_{M}^{ \ \ \A}
                                   + \frac{i}{2}  \EPS_{M}^{ \ \ b}
                                   (\E \Sm)_{ \ \ \dot\A}^{\A}
                                   \bar\D^{\dot\A} \bar\LAB
                                 \right )
\nonumber\\
\EPS_{M \ \dot\A} &\rightarrow& e^{2\LAB - \bar\LAB} \
                                 \left (
                                   \EPS_{M \ \dot\A}
                                   + \frac{i}{2}  \EPS_{M}^{ \ \ b}
                                   (\E \bar\Sm)_{\dot\A}^{ \ \ \A}
                                   \bar\D_{\A} \LAB
                                 \right )
\nonumber\\
\end{eqnarray}

In the end one can introduce a Weyl weight $\W_{w}$ and
the super-Weyl transformation law [\ref{wess_and_bagger}]

\begin{eqnarray}
\label{weight_rule}
    X^{\prime}  =  X   \ e^{- \W_{w}(\LAB + \bar\LAB)}
\nonumber \\
\end{eqnarray}

For the superfields of interest one finds

\begin{eqnarray}
\W_{w}(\EPS) = -2
\hspace{0.5cm}
,
\hspace{0.5cm}
 \W_{w}(\Si) = 0
\hspace{0.5cm}
 \mbox{and}
\hspace{0.5cm}
 \W_{w}(\OA) = \W_{w}(l) = \W_{w}(L) = 2.
\end{eqnarray}

If one performs now a K\"ahler transformation, the
lagrangian (\ref{lorentz_chiral_action}) is invariant under
the combined super-Weyl-K\"ahler transformation if
$F(\Si) = 6 \LAB$.
Note that $\lagrange_{pot}$ is independent of $S + \bar S$
if $\Om \neq \Om(S + \bar S)$.
Performing the same manipulations as in the $U_{K}(1)$-superspace
one finds the unconstrained lagrangian

\begin{eqnarray}
\label{u_lorentz_action}
\lagrange_{u} &=&
 \lagrange_{matter} + \lagrange_{YM} +
 \lagrange_{lm}
= - 3 \int \EPS
    \left \{
     e^{-K/3} - \frac{2n}{3} \ U \ L
    \right \}
\end{eqnarray}

and by variation with respect to $U$

\begin{eqnarray}
L &=& - \ \frac{1}{2n} \frac{\partial K}{\partial U} \  e^{-K/3},
\end{eqnarray}

which becomes the duality relation in Lorentz-superspace if
$ K(U) = -4n \ \mbox{ln} U $:

\begin{eqnarray}
\label{lorentz_duality_relation}
L &=& \frac{2}{U} \ e^{- K/3}
\end{eqnarray}

This duality relation looks different to (\ref{duality_relation})
and it seems, that the whole K\"ahler potential is involved.
But it only looks like, because we are not in an Einstein
normalized frame. That is to say, we still have to perform
the Weyl-rescaling of the vielbein and the appropriate
shift of the gravitino. This manipulation is different than the one
in the pure chiral theory of [\ref{wess_and_bagger},\ref{sugra_1},
\ref{kugo}]. Inserting (\ref{lorentz_duality_relation}) in
(\ref{u_lorentz_action}) one ends up with the dual theory.

\begin{eqnarray}
\label{dual_lorentz_action}
\lagrange &=&
 (4n - 3) \int \EPS \ e^{-{\tilde K}/3}
\\
\tilde K &=& \frac{3}{3-4n} (4n \ \mbox{ln} L - 4n \ \mbox{ln} 2
                             - 3 \ \mbox{ln} (T + \bar T))
\end{eqnarray}

Performing now a Weyl-rescaling of the vielbein
$e_{m}^{ \ a} \rightarrow e_{m}^{ \ a} \ e^{\sigma}$
one finds in the notation of
[\ref{sugra_1}]

\begin{eqnarray}
\label{Weyl_rescaling_lorentz}
e^{2\sigma} &=& - \frac{3}{\phi}
\hspace{1cm} \mbox{and} \hspace{1cm}
\phi = (4n - 3) e^{-{\tilde K}/3}
\end{eqnarray}

The non canonical Einstein term of
(\ref{dual_lorentz_action}) gets Einstein normalized
as usual

\begin{eqnarray}
\label{Weyl_rescaling_curvature}
\frac{1}{6} e \phi \R  &\rightarrow&
- \frac{1}{2} e \R - \frac{3}{4} e (\partial_{m} \mbox{ln} \phi)^{2}
+ \frac{3}{2} \partial_{m} (e g^{mn}\phi^{-1} \partial_{n} \phi)
\end{eqnarray}

The dual lagrangian (\ref{dual_lorentz_action}) is also invariant
under the combined super-Weyl-K\"ahler symmetry,
although the linear multiplet $L$ enters
the K\"ahler potential now.
Again the super-Weyl-K\"ahler constraint reads
$F(\Si) = 6 \LAB$, but this time
the dual (ill-defined) K\"ahler potential $\tilde K$
transforms under super-Weyl transformation
as

\begin{eqnarray}
 \tilde K(\Si,\bar \Si)  \rightarrow  \tilde K(\Si,\bar \Si)
                          - \frac{24n}{3-4n} \ (\LAB + \bar\LAB)
\end{eqnarray}

Still the same arguments given in the
$U_{K}(1)$-superspace formulation yield a
well defined K\"ahler potential.


\section{Conclusions}
\resetcounter

It has been shown, that the three basic functions relevant for
model building in local effective string theories
in 4 dimensions, namely the K\"ahler potential,
the superpotential and the gauge coupling function
can be reduced in the linear representation
of the dilaton to two functions at tree level.
However one ends up again with three functions
if one identifies a well-defined K\"ahler potential
satisfying the K\"ahler condition.
Gaugino condensation has been studied
in the framework of $U_{K}(1)$-superspace
in the chiral and
the linear represention of the dilaton and it turns out, that the
known results of the chiral representation
are completely equivalent to the results in
the linear representation. For the case of one gaugino
condensate the supersymmetric breaking
procedure is not affected by the duality transformation.
Furthermore the duality transformation has been
analysed also in Lorentz-superspace and it has been
shown, that the field dependent Weyl-rescaling
differs from the known pure chiral case.
So in the framework of a superconformal theory
the gauging of the compensators
[\ref{sugra_1}, \ref{kugo}], which leads to
the Poincare group as the structure group only,
also differs
from the gauging in the chiral representation of the dilaton.


\hspace{0.5cm}

{\bf Acknowledgement:} We would like to thank
G. Lopes-Cardoso for many conversations and
B. Ovrut and S. Ferrara for helpful discussions.

\hspace{0.5cm}

%
%

\section*{References}
\begin{enumerate}
\item
\label{linear}
S. Ferrara, J. Wess and B. Zumino, Phys. Lett. {\bf B51} (1974) 239;
\\
W. Siegel, Phys. Lett. {\bf B85} (1979) 333;
\\
S. Ferrara and M. Villasante, Phys. Lett.{\bf  B186} (1986) 85;
\\
S. Cecotti, S. Ferrara and L. Girardello, Phys. Lett. {\bf B198} (1987) 336;
\\
B. Ovrut, Phys. Lett.{\bf  B205} (1988) 455;
\\
B. Ovrut and S.K. Rama, Nucl. Phys. {\bf B333} (1990) 380,
Phys. Lett. {\bf B254} (1991) 138;
\\
P. Binetruy,G. Girardi and R. Grimm, Phys. Lett. {\bf B265} (1991) 111;
\\
\item
\label{Grimm1}
P.~Binetruy, G.~Girardi, R.~Grimm and M.~M\"uller,
Phys. Lett. {\bf B195} (1987) 389;
\\
P.~Adamietz,P.~Binetruy, G.~Girardi and R.~Grimm,
Nucl. Phys. {\bf B401} (1993) 275.
\item
\label{Chern_Simons}
S.~Cecotti, S.~Ferrara and M.Villasante, Int.J.Mod.Phys. A2 (1987)
1839
\\
G.~Girardi, R.~Grimm,
Nucl.Phys. B292 (1987) 181.
\item
\label{ovrut_schwiebert}
B.A.~Ovrut, C.~Schwiebert, Nucl.Phys. {\bf B321} (1989) 163.
\item
\label{Cardoso}
G.~Lopes-Cardoso, B.~Ovrut, Nucl.Phys.{\bf B369} (1992) 351,
Nucl.Phys. {\bf B392} (1993) 315, Nucl.Phys. {\bf B418} (1994) 535.
\item
\label{derendinger}
J.P.~Derendinger, S.~Ferrara, C.~Kounnas, F.~Zwirner,
Nucl. Phys. {\bf B372}  (1992) 145.
\item
\label{derque}
J.P.~Derendinger, F.~Quevedo, M.~Quiros,
Nucl. Phys. {\bf B428} (1994) 282.
\item
\label{gaugino}
H.P. Nilles, Phys. Lett. {\bf B115} (1982) 193;
\\
S. Ferrara, L. Girardello and H.P. Nilles, Phys. Lett. {\bf B125} (1983) 457;
\\
J.P. Derendinger, L.E. Ibanez and H.P. Nilles, Phys. Lett. {\bf 155} (1985) 65;
\\
M. Dine, R. Rohm, N. Seiberg and E. Witten, Phys. Lett. {\bf 156} (1985) 55.
\item
\label{dual}
C.~Kounnas and M.~Porrati, Phys. Lett. {\bf B191} (1987) 91;
\\
N.V. Krasnikov, Phys. Lett. {\bf 193} (1987) 37;
\\
L. Dixon, in Proc. 15th APS D.P.F. Meeting, 1990;
J.A. Casas, Z. Lalak, C. Munoz and G.G. Ross, Nucl. Phys. {\bf 347} (1990) 243;
\\
T.R. Taylor, Phys. Lett. {\bf 252} (1990) 59;
\\
H.-P.~Nilles and M.~Olechowski, Phys. Lett. {\bf B248} (1990) 268;
\\
D. L\"ust and T. Taylor, Phys. Lett. {\bf B253} 91991) 335;
\\
M. Cvetic, A. Font, L.E. Ibanez, D. L\"ust and F. Quevedo, Nucl. Phys.
{\bf B361} (1991) 194;
\\
D. L\"ust and C. Munoz, Phys. Lett. {\bf B279} (1992) 272.
\item
\label{filq}
A.~Font, L.E.~Ib\`a\~nez, D.~L\"ust and F.~Quevedo, Phys. Lett.
{\bf B245} (1990) 401.
\item
\label{fmtv}
S.~Ferrara, N.~Magnoli, T.R.~Taylor and G.~Veneziano, Phys.
Lett. {\bf B245} (1990) 409.
\item
\label{witten_index}
E.~Witten, Nucl. Phys. {\bf B202} (1982) 253.
\item
\label{wess_and_bagger}
J.~Wess and J.~Bagger, Supersymmetry and Supergravity,
Princeton University.
\item
\label{Grimm_2}
P.~Binetruy, G.~Girardi, R.~Grimm and M.~M\"uller,
Phys.Lett. {\bf B189} (1987) 83;
\\
P.~Binetruy, G.~Girardi and R.~Grimm, LAPP-preprint LAPP-TH-275/90  (1990).
\item
\label{zumino}
B.~Zumino, Phys. Lett. {\bf B87} (1979) 203.
\item
\label{dim_red}
E.~Witten, Phys. Lett. {\bf B155} (1985) 151;
\\
S.~Ferrara, C.~Kounnas and M.~Porrati, Phys. Lett. {\bf B181} (1986) 263.
\item
\label{gpr}
A. Giveon, M. Porrati and E. Rabinovici, hep-th 9401139.
\item
\label{FLST}
S.~Ferrara, D.~L\"ust, A.~Shapere and S.~Theisen, Phys. Lett. {\bf B225}
(1989) 363.
\item
\label{martin_mueller}
M.~M\"uller, Nucl. Phys. {\bf B264}  (1986) 292.
\item
\label{sugra_1}
E.~Cremmer, S.~Ferrara, L.~Girardello and A.~Van~Proeyen,
Nucl. Phys. {\bf B212} (1983) 413;
\\
E.~Cremmer, B.~Julia, J.~Scherk, S.~Ferrara, L.~Girardello
and P.~Van~Nieuwenhuizen, Nucl. Phys. {\bf B147} (1979) 105.
\item
\label{taylor1}
G.~Veneziano, S.~Yankielowicz, Phys. Lett. {\bf B113} (1982) 231;
\\
T.R.~Taylor, G.~Veneziano, S.~Yankielowicz, Nucl. Phys. {\bf B218} (1983) 493.
\item
\label{DKL}
L. Dixon, V. Kaplunovsky and J. Louis, Nucl. Phys. {\bf B355} (1991) 649.
\item
\label{howe_tucker}
P.~Howe and R.~Tucker, Phys. Lett. {\bf B80}  (1978) 138.
\item
\label{kugo}
T.~Kugo and S.~Uehara, Nucl. Phys. {\bf B222} (1983) 125,
                       Nucl. Phys. {\bf B226}  (1983) 49.
\end{enumerate}

\end{document}